# Comparison of the effectiveness and costs of hepatitis A vaccination strategies to mitigate future deaths in the Republic of Korea

**Yuna Lim[1], Yubin Seo[2], Jacob Lee[2], and Eunok Jung[1]\***

[1] Department of Mathematics, Konkuk University, Seoul, Republic of Korea; [2] Division of Infectious Diseases, Department of Internal Medicine, Kangnam Sacred Heart Hospital, Hallym University College of Medicine, Seoul, Republic of Korea

*Background:* Improved hygiene and infant vaccinations have led to age-specific variations in hepatitis A antibody prevalence in Korea, with lower rates among individuals in their 20s–40s. Given that the fatality rate of hepatitis A increases for those aged 50 and older, the low immunity level among younger adults indicates a future risk of increased deaths in older age groups without additional preventive measures.

*Methods:* We developed an age-structured transmission model to assess the impact of adult vaccination, assuming it begins in 2025. The 20s age group was modeled with an additional compartment to account for hepatitis A vaccination administered to military recruits. Vaccination strategies targeting the 20s–30s and 40s–50s age groups were compared, considering antibody testing costs for the latter in Korea and focusing on projected deaths over approximately 50 years.

*Results:* When the total vaccination cost is fixed, targeting the 40s–50s group covers 20% fewer individuals than the 20s–30s group but yields a 1.3- to 1.5-fold greater reduction in deaths. When the total vaccine supply is fixed, targeting the 40s–50s group is 1.2 times more expensive but yields a 1.7- to 1.8-fold greater reduction in deaths than the 20s–30s group. Moreover, including a second dose for military personnel in the 20s–30s group has minimal impact on reducing deaths.

*Conclusions:* Preventive measures for adults may be necessary to reduce future hepatitis A-related deaths. Vaccinating the 40s–50s group would be more effective in reducing deaths than vaccinating the 20s–30s group.

# INTRODUCTION

Hepatitis A is a common viral illness worldwide, primarily transmitted through contaminated food, water, or personal contact. Before the 1980s, hepatitis A was widespread in the Republic of Korea, with many people contracting the virus during childhood without showing symptoms and subsequently developed lifelong immunity.[1] Over the past 50 years, socioeconomic development and improvements in public health and hygiene have led to a decline in childhood hepatitis A infections.[1] In Korea, since May 2015, hepatitis A vaccination has been included in the National Immunization Program for children aged 12 to 23 months, effectively reducing the risk of infection in children.[2] Consequently, the prevalence of hepatitis A antibodies now varies by age group: high in those under 20 due to vaccination and among adults over 50 owing to natural infection, whereas low in middle-aged adults—who had limited exposure to the virus and were not vaccinated.[2]

The fatality rate of hepatitis A increases with age: although it is 0.2% across all age groups, it rises to 1.8% among those aged 50 and older.[3,4] Without preventive measures for adults with low antibody prevalence, the low immunity levels among young adults today may contribute to increased risk of death in older age groups in the future.

No specific treatment exists for hepatitis A, making prevention crucial.[3] In Korea, the hepatitis A vaccination program has been actively implemented for infants and military recruits. However, for the general adult population, vaccination policy remains at the recommendation level. Because of cost-related challenges, vaccinating all individuals without antibodies may be infeasible. Therefore, to reduce future hepatitis A-related deaths, cost-effective and efficient adult vaccination strategies are essential.

Mathematical modeling in infectious diseases has played an important role in epidemic forecasting and in proposing effective intervention policies. Using a mathematical model that considers age and region, Van Effelterre *et al*. suggested that nationwide infant vaccination in the United States was most effective in reducing the incidence of hepatitis A.[4] Patterson *et al*. evaluated the impact and cost-effectiveness of various childhood vaccination scenarios on hepatitis A cases and deaths in South Africa.[5] Lin *et al*. demonstrated that an early hepatitis A vaccination campaign targeting men who had sex with men with HIV effectively reduced a hepatitis A outbreak in Taiwan from 2015 to 2017 using a transmission model.[6] Ayoun *et al*. proposed that in Tunisia, where the prevalence of hepatitis A was increasing in childhood and adolescence, a single dose of hepatitis A vaccine at birth followed by a booster at school entry was most effective in reducing incidence rates.[7]

Our study differs from previous research by developing a mathematical model to address potential future risks associated with age-specific hepatitis A fatality rates and uneven antibody prevalence across age groups in Korea. The model reveals that age groups with lower antibody prevalence are gradually shifting toward older age groups with higher fatality rates over time. As this shift occurs, infections in older age groups lead to more deaths, highlighting the need for proactive adult vaccination. Furthermore, this study aims to identify an effective strategy to reduce deaths by comparing vaccination approaches targeting the 20s–30s and 40s–50s age groups, both of which have low antibody prevalence.

# METHODS

## Age-structured hepatitis A model

The model we developed includes heterogeneity in the prevalence of hepatitis A antibodies and fatality rates across different age groups. The epidemiology of infection follows an SEIAQR deterministic model, where the subscript $i$ for each compartment denotes the age group. Compartments for vaccinated individuals and deaths are also included.

The population is divided into eight age groups based on the prevalence of antibodies: 0–9 ($G_0$), 10–19 ($G_1$), 20–29 ($G_2$), 30–39 ($G_3$), 40–49 ($G_4$), 50–59 ($G_5$), 60–69 ($G_6$), and 70 and older ($G_7$). The model flowchart is depicted in Figure 1. Susceptible ($S$) and vaccinated individuals ($V$) are exposed to the hepatitis A virus with a force of infection $\lambda(t)$ and become infectious ($I, A$) after $1/\kappa$ days. Here, 70% of exposed individuals develop jaundice symptoms ($I$), while 30% remain asymptomatic ($A$).[8,9] Asymptomatic individuals recover after $1/\eta$ days, while symptomatic individuals are assumed to be isolated ($Q$) immediately upon onset and then either recover ($R$) or die ($D$) after isolation. All age groups follow a generalized flowchart (Figure 1), with $G_2$ including an additional compartment ($V_M$) for individuals who received a single dose of vaccination in the military. Among them, those who complete the second dose move to $R_2$, while the rest either transition to the susceptible group of the next age group ($S_3$) after the duration of effectiveness of the single dose ($1/w$) or become infected and move to $E_2$. The system of equations, along with description and the values of the model parameters, can be found in the Supplementary Materials.

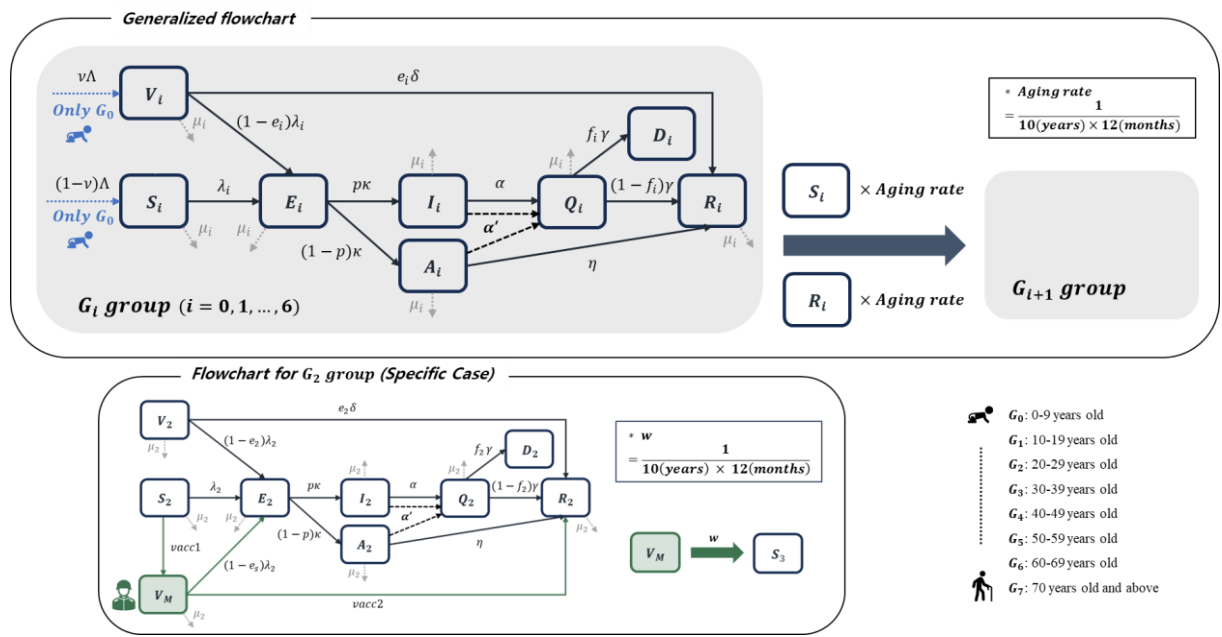

**Figure 1.** Model flowchart describing the transmission of hepatitis A. The susceptible class $S_i$ of age group $i$ can be exposed to hepatitis A virus ($E_i$) with the force of infection $\lambda_i$. Births $\Lambda$ are introduced only into the $S_0$ and $V_0$ of the group $G_0$ depending on the infant national vaccination rate $\nu$. According to the vaccine effectiveness of the first dose $e_i$, individuals in $V_i$ may either become fully protected from infection after the interval between the first and second doses $1/\rho$ or may be exposed to the hepatitis A virus with the force of infection. The mean latent period is $1/\kappa$, after which those in the exposed class $E_i$ move to $I_i$ or $A_i$ depending on the proportion $p$ of individuals who develop jaundice symptoms. The mean infectious period for asymptomatic individuals is $1/\eta$. Individuals in $I_i$ are isolated ($Q_i$) upon the onset of jaundice symptoms occurring after $1/\alpha$ days. Those in the isolated compartment die ($D_i$) or recover ($R_i$) after an average of $1/\gamma$ days. The parameters $f_i$ and $\mu_i$ represent the fatality rate owing to the disease and natural death rate, respectively, and $\alpha'$ indicates the effect of intervention when an external force of infection occurs in 2019 and 2021.

We consider aging in the model to investigate the number of deaths caused by hepatitis A over approximately 50 years. It takes 10 years for individuals in group $G_i$ to move to the next age group $G_{i+1}$, and since the model' time unit is monthly, the ageing rate is set to $1/(12 \times 10)$. Because the period from infection to recovery is faster than aging, the aging rate is applied only to the $S$ and $R$ compartments. The Supplementary Material demonstrates that the population structure produced by our model closely matches the future population projections by Statistics Korea.

The identified routes of hepatitis A infection in Korea include contact with infected individuals and consumption of contaminated water or food, such as salted clams which caused a relatively high number of cases in 2019 and 2021.[10–15] Specifically, approximately 17,600 cases occurred in 2019 and 6,600 in 2021. Furthermore, in 40–50% of cases, the source of infection remains unknown.[16] These infection routes are incorporated into the force of infection in our model.

The age-specific force of infection $\lambda_i$ is defined as

$$\lambda_i = \begin{cases} P\sum_{j=0}^{7}(C_1^{ij} + C_2^{ij} + C_3^{ij} + C_4^{ij})\frac{(I_j + A_j)}{N} + \sigma_i \psi_k + \omega_i \phi_1, & t_1 \leq t \leq t_2, \\ P\sum_{j=0}^{7}(C_1^{ij} + C_2^{ij} + C_3^{ij} + C_4^{ij})\frac{(I_j + A_j)}{N} + \sigma_i \psi_k + \omega_i \phi_2, & t_3 \leq t \leq t_4, \\ P\sum_{j=0}^{7}(C_1^{ij} + C_2^{ij} + C_3^{ij} + C_4^{ij})\frac{(I_j + A_j)}{N} + \sigma_i \psi_k, & \text{otherwise,} \end{cases}$$

where $N = \sum_{i=0}^{7}(S_i + V_i + E_i + I_i + A_i + R_i)$.

The first term denotes transmission through contact; the second term represents a persistent external force of infection, including contaminated clams, water, and unidentified sources; and the third term captures an external force of infection associated with specific salted clam events occurring from January ($t_1$) to August ($t_2$) in 2019 and January ($t_3$) to August ($t_4$) in 2021.

The parameter $P$ represents the probability of infection through contact, and $C_1$, $C_2$, $C_3$, and $C_4$ denote the average number of contacts per month from age group $i$ to $j$ in households, schools, workplaces, and other places, respectively.[15] Additionally, $\sigma_i$ indicates the age-dependent relative susceptibility to a persistent external force of infection compared to those aged 70 and older (i.e., $\sigma_7 = 1$), and $\psi_k$ represents the persistent external force of infection by year, where the subscript $k$ corresponds to the years 2016 to 2024. The parameter $\omega_i$ indicates the age-dependent relative intake rate of salted clams provided at restaurants compared to age 40–49 (i.e., $\omega_4 = 1$), and $\phi_1$ and $\phi_2$ denote the external forces of infection in 2019 and 2021, respectively. Additional details regarding the setting of the force of infection $\lambda_i$ are provided in the Supplementary Materials.

The unknown parameters $P$, $\sigma_i$, $\varphi_k$, $\omega_i$, $\phi_1$, $\phi_2$, and $\alpha'$ are estimated using the least square fitting method from hepatitis A case data provided by the Korea Disease Control and Prevention Agency (KDCA). Here, $\alpha' \in \{\alpha_1', \alpha_2'\}$ represents the effect of intervention when the external force of infection occurs in 2019 and 2021, respectively. See the Supplementary Materials for a detailed explanation of the parameter estimation and results. In this study, age-specific antibody prevalence is incorporated into the model through initial values set for $S_i$ and $R_i$.[17,18]

**Scenarios**

Extending the simulation based on the parameter estimation derived from data spanning 2016 to 2024, we investigate the expected number of hepatitis A-related deaths from 2025 to 2072, covering approximately 50 years. To compare the effects and costs of age-based vaccination strategies in reducing deaths, we design the following three scenarios:

- ○ ***S1.1:*** *Vaccination targeting the 20s–30s age group, including the second dose for individuals who received a single dose during the military service*

- *S1.2:* Vaccination targeting the 20s–30s age group, excluding the second dose for individuals who received a single dose during the military service
- *S2:* Vaccination targeting the 40s–50s age group

In all three scenarios, the number of vaccinations per month was fixed, with four levels considered: 5K, 100K, and 500K.

For comparison of the three scenarios, the following two criteria are fixed:

1. **Total cost:** *equivalent to the cost of administering two doses to 5M individuals*
2. **Total vaccine supply:** *equivalent to the amount required to administer two doses to 5M individuals (i.e., 10M doses)*

By fixing each criterion, we analyze the impact of the three scenarios on deaths reduction and associated costs. The 5M individuals represent approximately 10% of the population in the Republic of Korea and about 30% of those without hepatitis A antibodies. This value is determined based on model simulations to ensure that it does not exceed the number of individuals without antibodies in each age group (20s–30s and 40s–50s) as of 2025.

In Korea, individuals aged 40 and older are required to undergo an antibody test prior to vaccination, with only those without antibodies receiving the vaccine. In $S2$, the cost of antibody test must therefore be considered in addition to the vaccination cost. The formula for calculating the number of individuals requiring an antibody test is provided in the Supplementary Materials.

If an additional vaccination policy for adults is introduced, predicting the scale of government financial support is difficult. Therefore, we adopt a ratio-based cost estimation approach that does not rely on specific vaccine or test prices. In Korea, the cost of two doses of the hepatitis A vaccine ranges from 41.3 to 110 USD, and the cost of an antibody test ranges from 5.5 to 13.1 USD.[19–23] Based on these costs, we calculate the ratios of vaccination to antibody test costs. For the main scenario comparison, we select a cost ratio of 20:4 (vaccination cost to antibody test cost), which closely reflects the current situation in Korea. This ratio is based on the costs supported under the 2021 National Immunization Program for high-risk groups.[24–26] Results under other cost ratios are provided in the Supplementary Materials.

**RESULTS**

We first examined changes in age-specific antibody prevalence, as well as changes in the number of individuals without antibodies, cases, and deaths by age over approximately 50 years from 2025, assuming no additional preventive measures—such as adult vaccination—are implemented to prevent hepatitis A. Over time, antibody prevalence increases among individuals aged 40 years and younger, while it decreases among those aged 50 years and older (Figure 2a). In 2025, many susceptible individuals are present in $G_2$, $G_3$, and $G_4$; however, as younger individuals without antibodies age, the number of susceptible individuals gradually increases in older age groups over time (Figure 2b). Susceptibility to infection also rises with age (Supplementary Table 2). Consequently, the number of cases among the elderly increases over time (Figure 2c), with most annual hepatitis A-related deaths occurring in this population (Figure 2d). Because of both the high fatality rate and rising number of cases among the elderly, annual deaths also increase over time. The cumulative number of hepatitis A deaths is expected to reach 4,032 by 2072.

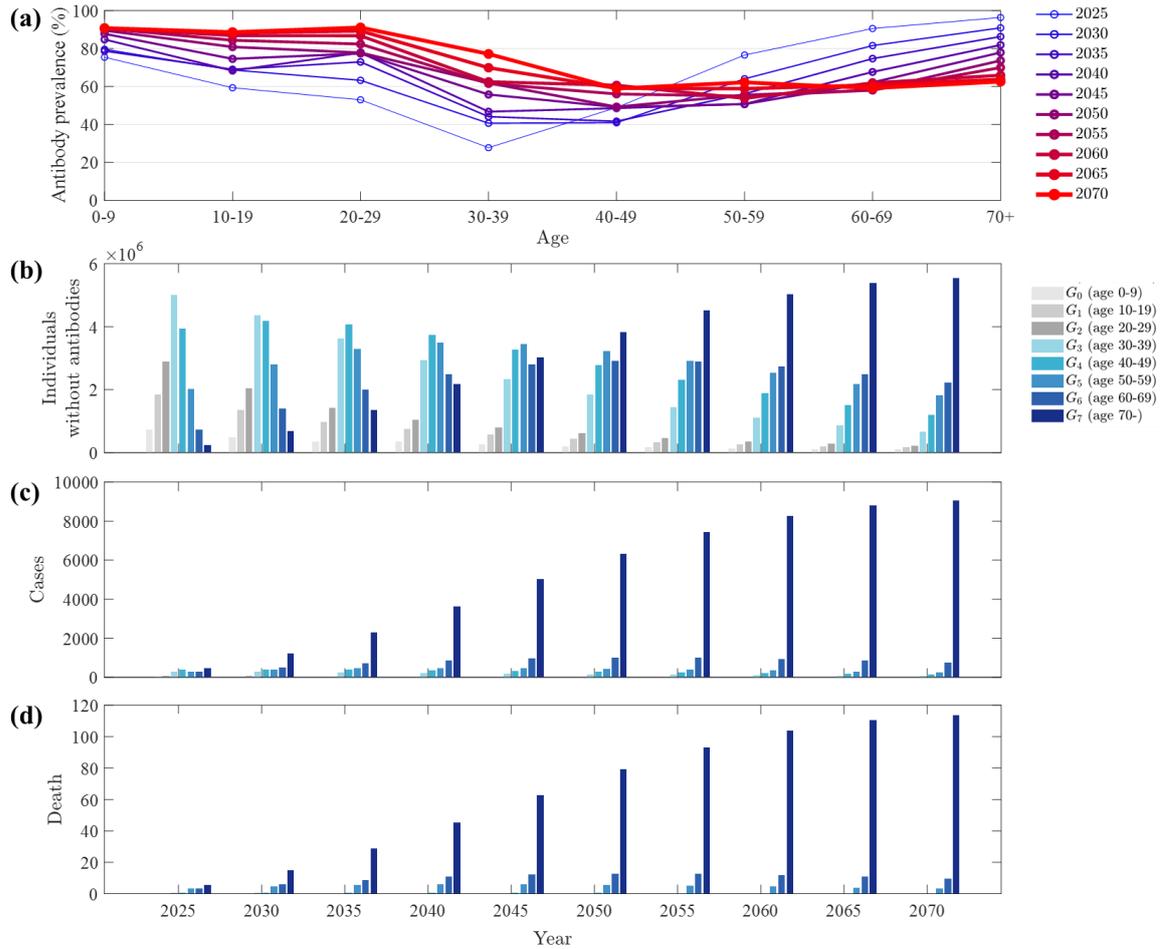

**Figure 2.** Projected changes in (a) age-specific antibody prevalence, (b) the number of individuals without antibodies, (c) the number of cases, and (d) annual deaths by age from 2025 to 2072, assuming no additional interventions targeting adults.

### 1. Fixed total cost: cost equivalent to administering two doses of vaccine to 5M individuals

When the total cost is fixed, we first examine how many people can receive vaccination and undergo an antibody test. Figure 3 shows the number of individuals eligible for vaccination in each scenario, assuming 50K individuals are vaccinated per month. In scenarios $S1.1$, $S1.2$, and $S2$, the number of individuals eligible for vaccination is 5,333,110, 5,000,000, and 3,926,030, respectively. For vaccination targeting the 20s–30s age group, two individuals who received a single dose in the military can be vaccinated for the cost of two doses for one civilian. Consequently, despite having the same total cost, the number of individuals eligible for vaccination differs between $S1.1$ and $S1.2$. For vaccination targeting the 40s–50s age group, the number of individuals eligible for vaccination differs between $S1.2$ and $S2$ because of the need to account for both vaccination and antibody test costs.

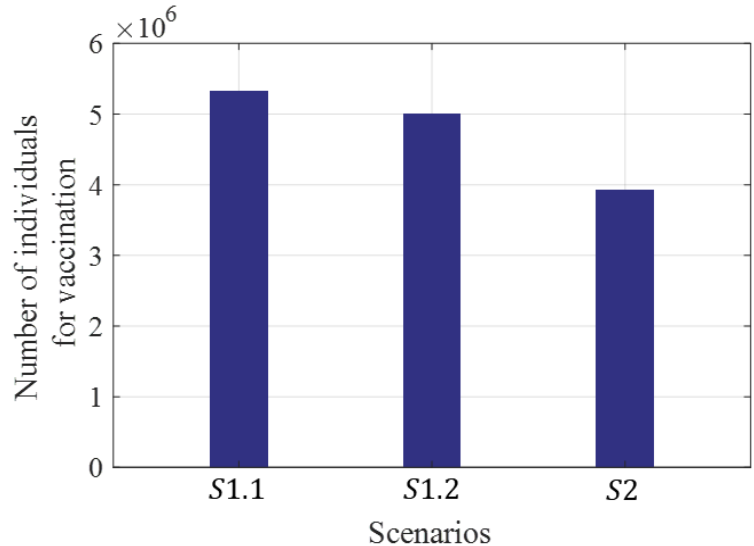

**Figure 3.** Number of individuals eligible for vaccination under a fixed total cost in each scenario—$S1.1$ (vaccination targeting the 20s–30s age group including second dose for military personnel), $S1.2$ (vaccination targeting the 20s–30s age group excluding second dose for military personnel), and $S2$ (vaccination targeting the 40s–50s age group)—assuming 50K individuals are vaccinated per month.

The number of individuals eligible for vaccination and antibody testing under various monthly vaccination counts is presented in Table 1. The number of individuals eligible for vaccination fluctuates depending on the number of vaccinations per month. This variation arises because, in $S1.1$, the number of eligible individuals is influenced by the proportion of individuals in the 20s–30s age group who received a single dose in the military, while in $S2$, it is affected by the proportion of individuals with and without antibodies, which determines the number of antibody tests required. Under a fixed total cost equivalent to administering two doses to 5M individuals, the number of individuals eligible for vaccination in $S2$ is at least 20% lower than in $S1.1$ and $S1.2$.

**Table 1. Number of individuals eligible for vaccination and antibody testing in each scenario by the number of vaccinations per month**

| Scenario | Number of individuals eligible for vaccination and antibody testing | | | | | |
|---|---|---|---|---|---|---|
| | Number of vaccinations per month | | | | | |
| | (50K) | | (100K) | | (500K) | |
| | Vaccination | Antibody test | Vaccination | Antibody test | Vaccination | Antibody test |
| $S1.1$ 20s–30s[1] | 5,333,110 | – | 5,324,290 | – | 5,355,420 | – |
| $S1.2$ 20s–30s[2] | 5,000,000 | – | 5,000,000 | – | 5,000,000 | – |
| $S2$ 40s–50s | 3,926,030 | 5,369,850 | 3,894,340 | 5,528,320 | 3,856,610 | 5,716,930 |

\* The second dose for military personnel refers to administering an additional dose to those who received a single dose during military service.
[1] Including the second dose for military personnel*
[2] Excluding the second dose for military personnel*

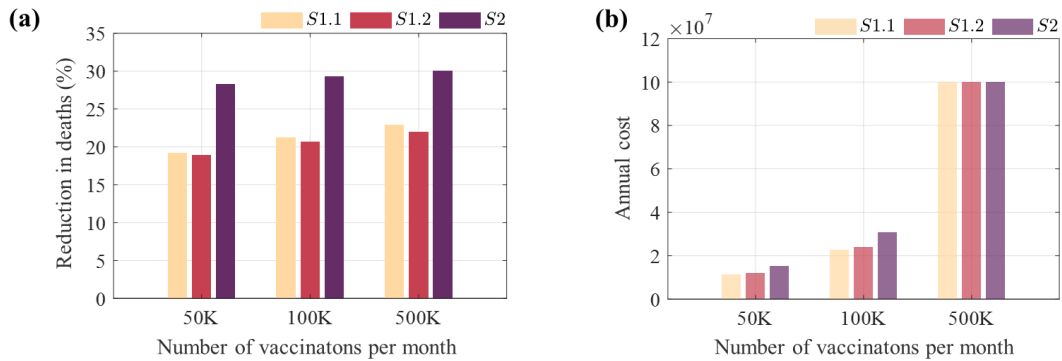

**Figure 4.** (a) Death reduction rates and (b) annual costs in each scenario by the number of vaccinations per month assuming a fixed total cost equivalent to administering two doses to 5M individuals.

Figure 4 presents the death reduction rates (panel a) and annual costs (panel b) in each scenario, based on the number of vaccinations per month as listed in Table 1. Here, the death reduction rate refers to the percentage reduction in deaths for each scenario, relative to the number of deaths expected over approximately 50 years in the absence of additional vaccination interventions for adults. The death reduction rates in $S2$ range from 28.3% to 29.7%, which are 1.3–1.5 times higher than those in $S1.1$ and $S1.2$. When vaccinating 50K or 100K individuals per month, the annual costs in $S2$ are 15M and 3M, respectively—approximately 1.3 times higher than those in $S1.1$ and $S1.2$. However, when vaccinating 500K individuals per month, the vaccination is completed within a year, resulting in equal annual costs of 100M across all scenarios.

For the vaccination targeting the 20s–30s age group, whether or not to include the second dose for individuals who received a single dose in the military makes no significant difference in either death reduction or annual cost. Additionally, while increasing the number of vaccinations per month—i.e., accelerating the vaccination—leads to higher annual costs across all scenarios, it provides only minimal reduction in deaths. See Supplementary Tables 5 and 6 for a summary of the death reduction rates and annual costs for each scenario by the number of vaccinations per month, along with results under various cost ratios.

### 2. Fixed total vaccine supply: amount of vaccine required for administering two doses to 5M individuals

We investigated the total cost, annual cost, and death reduction rate for each of the three scenarios under a fixed total vaccine supply of 10M doses. Figure 5 illustrates these outcomes based on the number of vaccinations per month. When the total vaccine supply is fixed, the total and annual costs in $S2$ are approximately 1.2 times higher than those in $S1.1$ and $S1.2$, regardless of the number of vaccinations per month. In contrast, the death reduction rates in $S2$ are approximately 1.7–1.8 times greater than in $S1.1$ and $S1.2$.

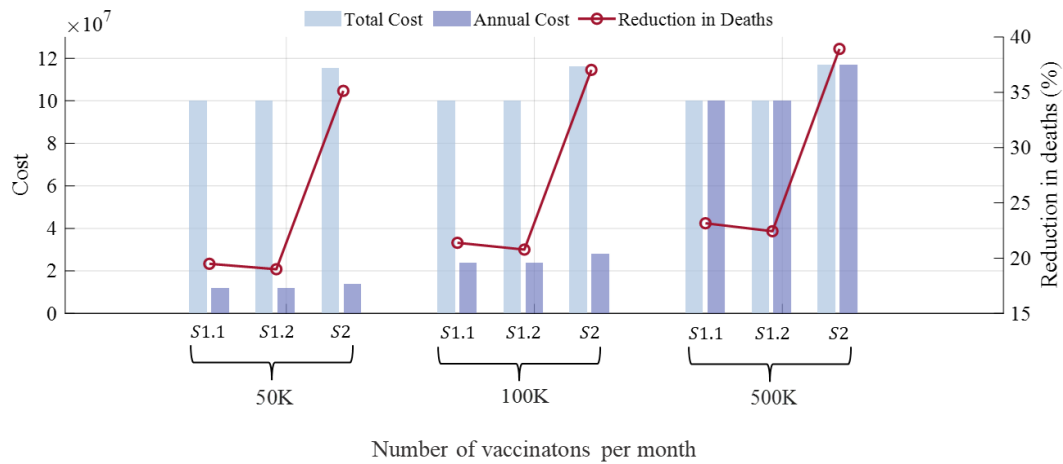

**Figure 5.** Total costs, annual costs, and reduction in deaths (death reduction rates) for each scenario by the number of vaccinations per month, assuming a fixed total vaccine supply of 10M doses

For vaccination targeting the 20–30s age group, the total cost in $S1.1$ is the same as in $S1.2$, and the annual cost is also identical owing to the fixed total vaccine supply, and the differences in death reduction rates between the two are negligible. Under a fixed total vaccine supply, increasing the number of vaccinations per month has little impact on reducing deaths but results in higher annual costs across all three scenarios. See Supplementary Figure 4 for a summary of total costs, annual costs, and death reduction rates for each scenario based on the number of vaccinations per month, along with results incorporating various cost ratios.

## DISCUSSION

The transition of individuals without hepatitis A antibodies from younger to older age groups may lead to a substantial increase in deaths in the future owing to the higher fatality rate associated with age. This study employed mathematical modeling to evaluate effective vaccination strategies for hepatitis A in the context of Korea's current epidemiological situation. If no additional hepatitis A vaccination is provided for adults and the current trend continues, the annual number of deaths is expected to rise, resulting in approximately 4,032 hepatitis A-related deaths by 2072. This result suggests that vaccination intervention for adults without antibodies may be necessary to reduce future deaths from hepatitis A. We modeled this phenomenon in Korea mathematically and present the following findings.

Our study shows that, under a fixed total cost, vaccinating the 40s–50s age group covers 0.2 times fewer individuals than vaccinating the 20s–30s age group but results in 1.3–1.5 times greater reduction in deaths. The annual cost of vaccinating the 40s–50s age group is up to approximately 1.3 times higher—or similar to that of vaccinating the 20s-30s age group—depending on the number of vaccinations per month. When the total vaccine supply is fixed, the total and annual costs for vaccinating the 40s–50s age group are about 1.2 times higher than those for the 20s–30s age group, yet the reduction in deaths is 1.7–1.8 times greater. Under both criteria, including or excluding the second dose for individuals who received a single dose during military service does not lead to significant differences in the reduction in deaths when targeting the 20s–30s age group. This suggests that, if an additional adult vaccination program is introduced, military personnel may continue to receive only a single dose as is current practice. Finally, we found that increasing the number of vaccinations per month raises annual costs without significantly improving the reduction in deaths.

The limitations of this study are as follows. Firstly, this study did not consider the economic losses caused by restricted labor participation owing to hepatitis A infection in the 20s–30s age group, which has the highest incidence in Korea. Instead, it focused solely on reducing deaths. Therefore, an additional cost-effectiveness analysis that includes social costs, such as deaths and years of life lost due to infection, is needed. Secondly, in Korea, adults can receive hepatitis A vaccination voluntarily. Furthermore, a significant hepatitis A outbreak related to contaminated clams occurred in Korea in 2019, which prompted the government to recommend vaccination for adults. Aside from records of vaccinations for military recruits and the hepatitis A vaccination support program for high-risk groups (e.g., patients with chronic liver disease) implemented from 2020 to 2021, no comprehensive record of adult hepatitis A vaccinations exists. Since we did not account for all vaccinated adults in our model, the number of unvaccinated adults may be overestimated. Lastly, we predicted future deaths from 2025 onward by fixing the value of $\psi_{2024}$ estimated for the year 2024. However, this parameter may vary due to nationwide interventions, such as improvements in groundwater quality, which could lead to discrepancies between our predictions and actual future outcomes.

## CONCLUSIONS

In summary, without additional adult vaccination, the risk of deaths in older age groups may increase in the future. If vaccination is introduced for adults in their 20s–30s, whether or not a second dose is included for those who received a single dose in the military makes little difference in the reduction of deaths. Vaccination targeting the 40s–50s age group is more cost-effective in reducing deaths than targeting the 20s–30s age group when the cost is equal, and results in a greater reduction in deaths when the vaccine supply is the same.

## DECLARATIONS

### ETHICS APPROVAL AND CONSENT TO PARTICIPATE

Not applicable

### CONSENT FOR PUBLICATION

Not applicable

### AVAILABILITY OF DATA AND MATERIALS

The datasets used in the current study are available from the Korea Disease Control and Prevention Agency (KDCA) website: https://www.kdca.go.kr/.

### COMPETING INTERESTS

The authors declare that they have no competing interests.

### FUNDING

This research was supported by 'The Government-wide R&D to Advance Infectious Disease Prevention and Control', Republic of Korea (grant number: HG23C1629; to E.J.). This paper was also supported by the Korea National Research Foundation (NRF) grant funded by the Korean government (MEST) (NRF-2021R1A2C100448711; to E.J.).

**AUTHOR'S CONTRIBUTIONS**

Y.L. contributed to the conceptualization, methodology, investigation, formal analysis, software, visualization, writing - original draft, writing - review & editing of the study. E.J. contributed to the conceptualization, formal analysis, funding acquisition, investigation, methodology, supervision, validation, writing - review & editing of the study. Y.S. and J.L. contributed to the conceptualization, investigation, validation, writing - review & editing of the study.

**ACKNOWLEDGEMENTS**

Not applicable

## SUPPLEMENTARY MATERIAL

**System of equations of the mathematical model of hepatitis A transmission in Korea**

The following system of ordinary differential equations describes the age-structured model used to simulate the transmission dynamics of hepatitis A in Korea:

$$\frac{dS_j}{dt} = -\lambda_j S_j - u_j S_j - \delta_{j2}(vacc1) + \delta_{j3}(wV_M) + IN_j^S - OUT_j^S,$$

$$\frac{dV_j}{dt} = -(1-e_j)\lambda_j V_j - e_j \rho V_j - u_j V_j + IN_j^V,$$

$$\frac{dE_j}{dt} = \lambda_j S_j + (1-e_j)\lambda_j V_j - \kappa E_j - \mu_j E_j + \delta_{j2}\left((1-e_s)\lambda_j V_M\right),$$

$$\frac{dI_j}{dt} = p\kappa E_j - \alpha I_j - \mu_j I_j - \alpha'(t) I_j,$$

$$\frac{dA_j}{dt} = (1-p)\kappa E_j - \eta A_j - \mu_j A_j - \alpha'(t) A_j,$$

$$\frac{dQ_j}{dt} = \alpha I_j - \gamma Q_j - \mu_j Q_j + \alpha'(t)(I_j + A_j),$$

$$\frac{dR_j}{dt} = (1-f_j)\gamma Q_j + \eta A_j + e_j \rho V_j - \mu_j R_j + \delta_{j2}(vacc2) + IN_j^R - OUT_j^R,$$

$$\frac{dD_j}{dt} = f_j \gamma Q_j,$$

$$\frac{dV_M}{dt} = \delta_{j2}(vacc1 - (1-e_s)\lambda_2 V_M - vacc2 - \mu_2 V_M - wV_M),$$

$$IN_j^S = \begin{cases} (1-v)\Lambda, & (j=0) \\ S_{j-1} \times Aging\ rate, & (j=1,2,3,4,5,6,7) \end{cases},$$

$$OUT_j^S = \begin{cases} S_j \times Aging\ rate, & (j=0,1,2,3,4,5,6) \\ 0, & (j=7) \end{cases},$$

$$IN_j^V = \begin{cases} v\Lambda, & (j=0) \\ 0, & (j=1,2,3,4,5,6,7) \end{cases},$$

$$IN_j^R = \begin{cases} 0, & (j=0) \\ R_{j-1} \times Aging\ rate, & (j=1,2,3,4,5,6,7) \end{cases},$$

$$OUT_j^R = \begin{cases} R_j \times Aging\ rate, & (j=0,1,2,3,4,5,6) \\ 0, & (j=7) \end{cases},$$

$$\alpha'(t) = \begin{cases} \alpha_1': Estimated\ constatnt\ value\ 1, & \tau_1 \leq t \leq \tau_2 \\ \alpha_2': Estimated\ constatnt\ value\ 2, & \tau_3 \leq t \leq \tau_4 \\ 0, & otherwise \end{cases},$$

where $j \in \{0, 1, 2, 3, 4, 5, 6, 7\}$, and $\tau_1, \tau_2, \tau_3$, and $\tau_4$ represent September 2019, October 2019, May 2021, and July 2021, respectively. The monthly number of births $\Lambda$ is derived from birth data and future population projection data provided by Statistics Korea.[1] In our model, only group $G_0$ receives vaccination owing to national vaccination program for infants. The number of military personnel receiving the first and second doses each month, denoted by $vacc1$ and $vacc2$, respectively, is derived from military vaccination data.[2] $\delta_{j2}$ is the Kronecker delta, which takes the value of 1 when $j = 2$ and 0 otherwise. The description and values of the model parameters are listed in Supplementary Table 1.

**Supplementary Table 1. Description and value of the model parameters**

| Symbol | Description | Value | Unit | Reference |
|---|---|---|---|---|
| $1/\kappa$ | Mean latent period | 14/30 | $month$ | (3,4) |
| $1/\alpha$ | Mean infectious period before icteric phase | 24/30 | $month$ | (3–5) |
| $1/\gamma$ | Isolation period | 7/30 | $month$ | (4) |
| $1/\eta$ | Mean infectious period | 31/30 | $month$ | (4,5) |
| $f_j$ | Fatality rate due to the hepatitis A by group ($j = 0, 1,…, 7$) | 0.002 ($j < 5$) <br> 0.018 ($j \geq 5$) | - | (3,6) |
| $v$ | Child vaccination rate | 0.95 | - | (7) |
| $e_j$ | Vaccine effectiveness of the first dose ($j = 0, 1,…7$) | 0.985 ($j = 0$) <br> 0.97 ($j \neq 0$) | - | (8,9) |
| $1/\rho$ | Interval between the first and the second dose | 360/30 | $month$ | (8) |
| $1/w$ | Duration of effectiveness of a single dose | 3600/30 | $month$ | (10,11) |
| $e_S$ | Effectiveness of single-dose administration in military personnel | 0.7585 | - | (12) |
| $p$ | Proportion of individuals isolated owing to symptom onset | 0.7 | - | (13) |
| $\mu_j$ | Natural fatality rate by group ($j = 0, 1,…,7$) | 0.0003 ($j = 0$) <br> 0.0001 ($j = 1$) <br> 0.0004 ($j = 2$) <br> 0.0007 ($j = 3$) <br> 0.0014 ($j = 4$) <br> 0.0033 ($j = 5$) <br> 0.0077 ($j = 6$) <br> 0.0352 ($j = 7$) | $month^{-1}$ | (14) |

**Setting of force of infection and results of parameters estimation**

Supplementary Figure 1 illustrates the results of parameter estimation. The shaded red areas indicate the periods when contaminated salted clam events occurred. The red squares represent the cumulative case data by age group from 2016 to 2024, with data trends for groups $G_6$ and $G_7$ differing from those of other age groups. This implies that these two age groups were rarely affected even during the events. However, considering the consistent occurrence of cases even among the older groups $G_6$ and $G_7$ with fewer contacts, the factor of persistent external force of infection needs to be included in the force of infection $\lambda_i$.

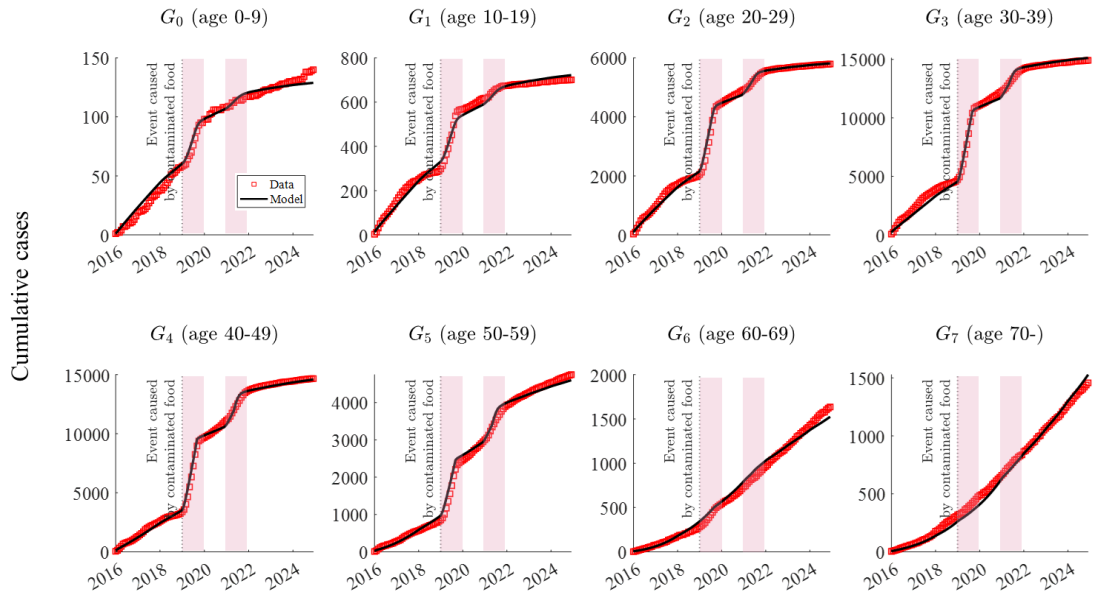

**Supplementary Figure 1.** Parameter estimation results using cumulative case data for eight age groups from 2016 to 2024.

Twenty-five parameters ($P, \sigma_i, \psi_k, \omega_i, \phi_1, \phi_2$) constituting $\lambda_i$, along with $\alpha_1'$ and $\alpha_2'$, are estimated using the least-square fitting method, minimizing the errors between the fitted model and the cumulative case data for eight age groups from January 2016 to December 2024.

To estimate the age-dependent relative susceptibility to persistent external force of infection, $\sigma_7$ of group $G_7$, which consistently had cases without being significantly affected by the events, is set to 1. To estimate the age-dependent relative intake rate of salted clam provided at restaurants, $\omega_4$ is set to 1 as group $G_4$ had the highest number of cases during the events. Also, purification of contaminated groundwater has been ongoing, and groundwater quality management projects have been pursued, so $\psi_k$ is estimated annually.

As $G_0$ and $G_1$ have relatively high contact with individuals, even with small values of the estimated parameters $\omega_0$ and $\omega_1$, the number of cases increases significantly during the events. The relatively small values of the estimated parameters $\omega_6$ and $\omega_7$ corresponded to the data indicating that $G_6$ and $G_7$ were not rarely affected during the events.

Supplementary Table 2 displays the number of individuals without antibody, cumulative cases, and ratios of cumulative cases to the number of individuals without antibodies, which increase with age. This confirms that the age-dependent relative susceptibility to persistent external force of infection $\sigma_i$ also increases with age. A description and value of the estimated parameters are listed in Supplementary Table 3.

**Supplementary Table 2.** Number of individuals without antibodies, cumulative cases, and ratios of cumulative cases to the number of individuals without antibodies by age group in 2016

| Group | Individuals without antibodies (a) | Cumulative cases in 2016 (b) | b/a (%) |
|---|---|---|---|
| $G_0$ | 1,587,582 | 16 | 0.001 |
| $G_1$ | 2,913,308 | 140 | 0.004 |
| $G_2$ | 6,454,068 | 889 | 0.013 |
| $G_3$ | 5,231,599 | 2,031 | 0.038 |
| $G_4$ | 1,745,222 | 1,209 | 0.069 |
| $G_5$ | 191,470 | 261 | 0.136 |
| $G_6$ | 15,220 | 72 | 0.473 |
| $G_7$ | 4,583 | 61 | 1.33 |

**Supplementary Table 3. Description and value of the estimated parameters**

| Symbol | Description | Value |
|---|---|---|
| $P$ | Probability that a contact with a susceptible individual results in transmission | 0.0048 |
| $\sigma_0$ | Relative susceptibility to persistent external force of infection compared to $G_7$ in $G_0$ | 0.0011 |
| $\sigma_1$ | Relative susceptibility to persistent external force of infection compared to $G_7$ in $G_1$ | 0.0035 |
| $\sigma_2$ | Relative susceptibility to persistent external force of infection compared to $G_7$ in $G_2$ | 0.0140 |
| $\sigma_3$ | Relative susceptibility to persistent external force of infection compared to $G_7$ in $G_3$ | 0.0321 |
| $\sigma_4$ | Relative susceptibility to persistent external force of infection compared to $G_7$ in $G_4$ | 0.0554 |
| $\sigma_5$ | Relative susceptibility to persistent external force of infection compared to $G_7$ in $G_5$ | 0.0791 |
| $\sigma_6$ | Relative susceptibility to persistent external force of infection compared to $G_7$ in $G_6$ | 0.2083 |
| $\psi_{2016}$ | Persistent external force of infection in 2016 | 0.00099 |
| $\psi_{2017}$ | Persistent external force of infection in 2017 | 0.00098 |
| $\psi_{2018}$ | Persistent external force of infection in 2018 | 0.00077 |
| $\psi_{2019}$ | Persistent external force of infection in 2019 | 0.0005 |
| $\psi_{2020}$ | Persistent external force of infection in 2020 | 0.00049 |
| $\psi_{2021}$ | Persistent external force of infection in 2021 | 0.00029 |
| $\psi_{2022}$ | Persistent external force of infection in 2022 | 0.00021 |
| $\psi_{2023}$ | Persistent external force of infection in 2023 | 0.00017 |
| $\psi_{2024}$ | Persistent external force of infection in 2024 | 0.00013 |
| $\omega_0$ | Relative intake rate of salted clams provided at restaurants compared to $G_4$ in $G_0$ | 0.0021 |
| $\omega_1$ | Relative intake rate of salted clams provided at restaurants compared to $G_4$ in $G_1$ | 0.0161 |
| $\omega_2$ | Relative intake rate of salted clams provided at restaurants compared to $G_4$ in $G_2$ | 0.2201 |
| $\omega_3$ | Relative intake rate of salted clams provided at restaurants compared to $G_4$ in $G_3$ | 0.5576 |
| $\omega_5$ | Relative intake rate of salted clams provided at restaurants compared to $G_4$ in $G_5$ | 0.8004 |
| $\omega_6$ | Relative intake rate of salted clams provided at restaurants compared to $G_4$ in $G_6$ | 0.1757 |
| $\omega_7$ | Relative intake rate of salted clams provided at restaurants compared to $G_4$ in $G_7$ | 0.0857 |
| $\phi_1$ | External force of infection in 2019 | 0.00034 |
| $\phi_2$ | External force of infection in 2021 | 0.00011 |
| $\alpha'_1$ | Effect of intervention when external force of infection occurs in 2019 | 1.5519 |
| $\alpha'_2$ | Effect of intervention when external force of infection occurs in 2021 | 1.4360 |

**Population projections**

Our model incorporates population movement between age groups owing to aging. Therefore, to assess whether the population structure generated by our model aligns with the future population structure, we compared it with the estimated population provided by Statistics Korea. Supplementary Figure 2 illustrates the population structures in 2025 (a) and 2070 (b) from our model. This represents the expected trends of low birth rates and aging in Korea. Supplementary Table 4 presents the total population, population aged 70 and above, and the proportion of population aged 70 and above in 2070, estimated by both our model and Statistics Korea. The values obtained by our model are similar to the estimates provided by Statistics Korea. This demonstrates that our model effectively reflects the future population structure.

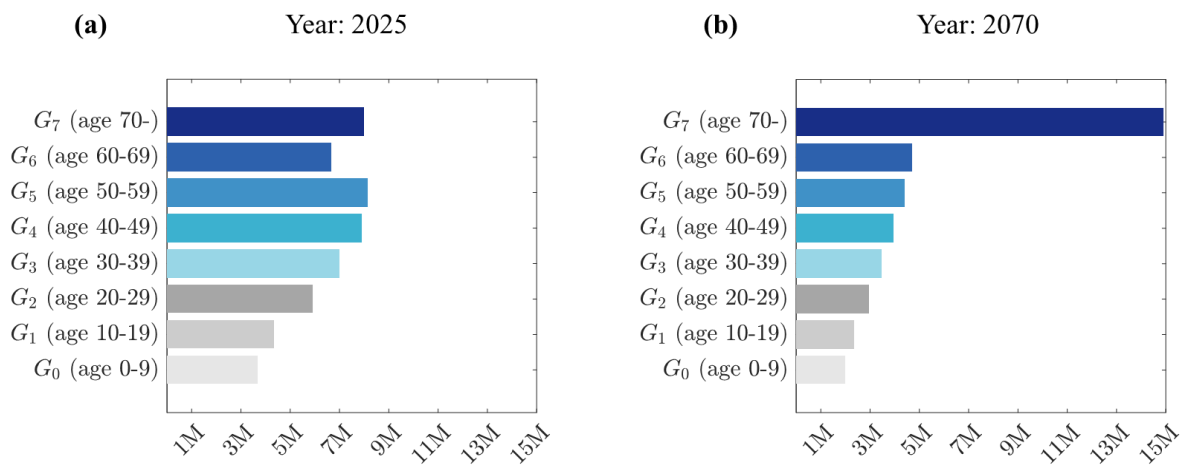

**Supplementary Figure 2.** Future population structures in 2024 (a) and 2070 (b) from the model.

**Supplementary Table 4. Population projections for the total population and those aged 70 and above, including their proportions, in 2070 estimated by our model and Statistics Korea**

| | Our model | Statistics Korea | Reference |
|---|---|---|---|
| Total population | 38,778,599 | About 37,660,000 | (16) |
| Population projection for those aged 70 and above | 14,922,000 | About 14,860,000 | (16) |
| Proportion of population aged 70 and above | 38.5% | 39.5% | |

**Formula for calculating the number of individuals requiring antibody testing**

We define the two variables to calculate the number of monthly antibody tests:

- $S_{old}$: individuals aged 40 and older who do not have antibodies
- $\tilde{R}$: Individuals aged 40 and older who were previously infected asymptomatically, recovered, and developed antibodies

The number of $\tilde{R}$ is calculated by multiplying the number of individuals aged 40 and older who have antibodies ($R_{old}$) by the proportion of individuals who develop jaundice symptoms with the rate of $p$. The individuals undergoing antibody testing are $S$ and $\tilde{R}$, as they either have never been infected or have recovered asymptomatically without knowing they were infected. Therefore, they will undergo antibody testing before vaccination.

The number of monthly antibody tests is derived from the following proportion:

(The number of vaccinations per month): (The number of antibody-tested individuals per month) =$S_{old}$: $(S_{old} + \tilde{R})$

Accordingly, the number of monthly antibody tests is calculated as follows. To explain the calculation process, we define the following variables:

- $t = 1, 2, \ldots, t_{end}$: time (unit: month)
- $T(t)$: number of antibody-tested individuals at time $t$
- $S_{old}(t)$: number of individuals without antibodies at time $t$
- $\tilde{R}_{nonT}(i)$: number of $\tilde{R}(t)$ individuals who have not undergone antibody test up to time $t$
- $\tilde{R}_T(i)$: number of $\tilde{R}(t)$ individuals who have undergone antibody test up to time $t$
- $v$: number of vaccinations per month

$$\tilde{R}_{nonT}(1) = \tilde{R}(1),$$

$$\tilde{R}_T(1) = v \times \frac{\tilde{R}_{nonT}(1)}{S(1)},$$

$$\tilde{R}_{nonT}(i) = \tilde{R}(i) - \tilde{R}_T(i-1),$$

$$\tilde{R}_T(i) = \tilde{R}_T(i-1) + v \times \frac{\tilde{R}_{nonT}(i)}{S(i)},$$

$$\boldsymbol{T}(i) = v \times \frac{S(i) + \tilde{\boldsymbol{R}}_{nonT}(i)}{S(i)}.$$

Here, $\tilde{R}_{nonT}(1)$ represents the number of antibody-positive individuals who do not undergo antibody testing at t = 1, while $\tilde{R}_T(1)$ represents those who do at t = 1. $\boldsymbol{\tilde{R}_{nonT}}(i)$ denotes the number of antibody-positive individuals who remain untested up to $t = i$ ($i \geq 1$), and $\tilde{R}_T(i)$ represents those who have undergone testing up to $t = i$. Finally, $\boldsymbol{T}(i)$ is the number of antibody-tested individuals at $t = i$.

**Results for various cost ratios of vaccination to antibody testing**

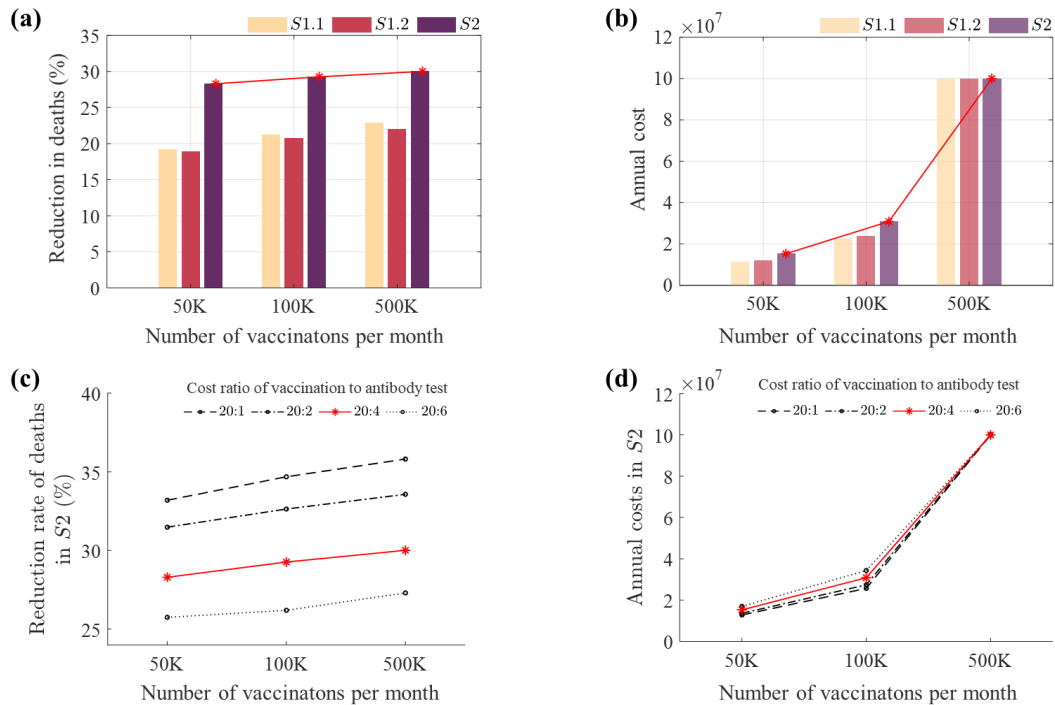

**Supplementary Figure 3.** (a) Reduction in deaths across the three scenarios, (b) annual costs across the three scenarios, (c) reduction in deaths in $S2$ based on the cost ratio of vaccination to antibody test, and annual costs in $S2$ based on the cost ratio of vaccination to antibody testing, and (d) all calculated according to the number of vaccinations per month, with the total cost fixed to the equivalent of administering two doses of vaccine to 5M individuals.

In panel (a), the red line represents the reduction in deaths in $S2$ when the cost ratio is 20:4, which aligns with the red line in panel (c). In panel (b), the red line represents the annual costs in $S2$ when the cost ratio is 20:4, which aligns with the red line in panel (d). As the cost ratio of vaccination to antibody testing decreases, the reduction in deaths in $S2$ may become greater. When the cost ratio is largest at 20:6, the reduction in deaths in $S2$ is the smallest, but it is still greater than in $S1.1$ and $S1.2$. Annual costs decrease as the cost ratio decreases, and increase as the cost ratio increases, although the differences are minimal. Supplementary Tables 5 and 6 summarize Supplementary Figure 3.

**Supplementary Table 5. Reduction in deaths across the three scenarios based on the cost ratio of vaccination to antibody testing and the number of vaccinations per month**

| Ratio of vaccination to antibody testing costs | Scenario | Reduction in deaths (%) | | |
|---|---|---|---|---|
| | | Number of vaccinations per month | | |
| | | (50K) | (100K) | (500K) |
| All ratios (20:1, 20:2, 20:4, 20:6) | *S*1.1 20s–30s (Including the second dose for military personnel*) | 19.2 | 21.2 | 22.9 |
| | *S*1.2 20s–30s (Excluding the second dose for military personnel*) | 18.9 | 20.7 | 22.0 |
| 20:1 | *S*2 40s–50s | 33.2 | 34.7 | 35.8 |
| 20:2 | | 31.4 | 32.6 | 33.6 |
| 20:4 | | 28.3 | 29.3 | 30.0 |
| 20:6 | | 25.8 | 26.2 | 27.3 |

**Supplementary Table 6. Annual costs in deaths across the three scenarios based on the cost ratio of vaccination to antibody testing and the number of vaccinations per month**

| Ratio of vaccination to antibody testing costs | Scenario | Annual cost | | |
|---|---|---|---|---|
| | | Number of vaccinations per month | | |
| | | (50K) | (100K) | (500K) |
| All ratios (20:1, 20:2, 20:4, 20:6) | *S*1.1 20s–30s (Including the second dose for military personnel*) | 11M | 23M | 100M |
| | *S*1.2 20s–30s (Excluding the second dose for military personnel*) | 12M | 24M | 100M |
| 20:1 | *S*2 40s–50s | 13M | 26M | 100M |
| 20:2 | | 14M | 27M | 100M |
| 20:4 | | 15M | 31M | 100M |
| 20:6 | | 17M | 34M | 100M |

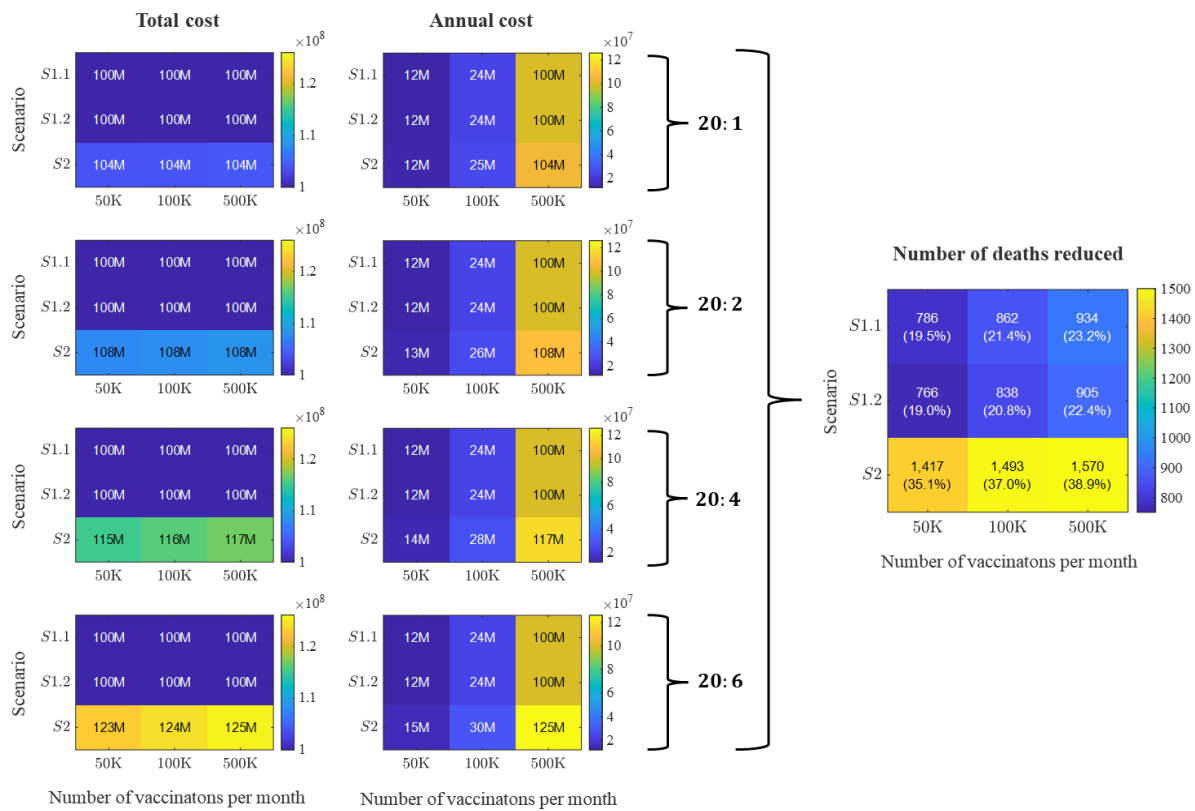

**Supplementary Figure 4.** Total costs, annual costs, and number of deaths reduced (death reduction rates) across the three scenarios based on the cost ratio of vaccination to antibody testing and the number of vaccinations per month when total vaccine supply is fixed to the amount required for administering two doses to 5M individuals.

In the graph on the left, the first column represents total costs, and the second column represents annual costs, with each row indicating an increase in the cost ratio of vaccination to antibody testing from top to bottom. S1.1 and S1.2 are unaffected by the cost ratio because the consideration of individuals undergoing antibody tests is unnecessary. With the total vaccine supply fixed, the total costs in these two scenarios remain constant, and only the annual costs increase as the number of vaccinations per month rises. In $S2$, both total costs and annual costs increase as the cost ratio increases. Since the total vaccine supply remains the same, the number of deaths reduced in each scenario is the same regardless of the cost ratio.